\documentclass{article}
\title{\bf Worldline deviations of charged spinning particles}
\author{M. Heydari-Fard$^1$, M. Mohseni$^2$\thanks{corresponding author. email:
m-mohseni@pnu.ac.ir} and H.R. Sepangi$^{1,3}$
\\ $^1${\small Department of Physics, Shahid Beheshti University, Evin, Tehran 19839, Iran}
\\$^2${\small Physics Department, Payame Noor University, Tehran 19395-4697,
Iran}\\$^3${\small Institute for Studies in Theoretical Physics
and Mathematics, Tehran, Iran}}
\begin{document}
\maketitle
\begin{abstract}
The geodesic deviation equation is generalized to worldline
deviation equations describing the relative accelerations of
charged spinning particles in the framework of Dixon-Souriau
equations of motion.


\end{abstract}
\section{Introduction}
In the study of the dynamics of particles in a given space-time,
one important object is the relative acceleration of particles.
For nearby test particles, this is given by the well-known
geodesic deviation equation. This equation gives in a
frame-independent way, how much two nearby geodesics deviate from
each other. For test particles which are not necessarily nearby,
similar equations may be derived by keeping higher order terms in
the approximation whose first order terms lead to the equation of
geodesic deviation. When the particles under study possess some
internal structure like spin, or if they are subject to extra
interactions like the Lorentz force, their worldlines will no
longer be geodesics. In such cases the geodesic deviation equation
should be modified so as to accommodate the effects of those extra
interactions on the relative accelerations of the particles. For
the case of test particles with charge moving in an arbitrary
space-time in the presence of some electromagnetic fields, a set
of worldline deviation equations has been obtained in \cite{bala}.
For particles with spin (but no charge), described by the
Mathison-Papapetrou-Dixon equations, a set of such generalized
worldline deviation equations was derived in a recent publication
\cite{moh2}. In \cite{chi} the relative motion of a spinning
particle with respect to a nearby free test particle in the
gravitational field of a rotating source was studied. The aim of
the present work is to obtain worldline deviation equations for
charged spinning particles moving in an arbitrary space-time in
the presence of electromagnetic fields.

In general relativity, the motion of charged spinning particles is
described by the so called Dixon-Souriau (DS, for abbreviation)
equations \cite{dixon2},\cite{souriau4}. These equations reflect
the effects of the spin-curvature, the charge-electromagnetic
field, and the spin-electromagnetic field on the motion of the
particle. These equations consist of seven independent equations
to describe the particle's four-momentum and spin tensor
supplemented by three other equations to render the equations of
motion complete. The particle's trajectory is then determined by
integrating its four-velocity which is obtained indirectly from
the equations of motion. These equations or simplified versions of
them have been used to study the motion of charged spinning
particles in different space-times \cite{hoj}-\cite{pas}. An
extension of these equations to the case where torsion fields are
also included was obtained in \cite{sol}, \cite{cog}. It has also
been shown that the DS equations reduce to the well known
Bargmann-Michel-Telegdi equations in the limit of the weak and
homogeneous external field \cite{zer}.

In the following sections we first review the DS equations briefly
and derive the worldline deviation equations along the lines of
the reference \cite{moh2}. We then apply these equations to the
case of motion of charged spinning particles in a gravitational
wave space-time and a uniform magnetic field and show how they can
be used to calculate the relative accelerations and also to
generate approximate solutions to the DS equations via a known
one. In the last section we present our conclusions.
\section{The DS equations of motion}
The motion of a charged spinning particle is described by the
Dixon-Souriau equations \cite{bai}
\begin{equation}
\frac{Dp^\mu}{D\tau}=-\frac{1}{2}{
R^\mu}_{\nu\lambda\rho}s^{\lambda\rho}{\dot
x}^\nu+q{F^\mu}_{\beta}{\dot x
}^\beta+\frac{k}{2}s^{\kappa\rho}D^\mu F_{\kappa\rho}, \label{eq1}
\end{equation}
\begin{equation}
\frac{Ds^{\mu\nu}}{D\tau}=p^\mu{\dot x}^\nu-p^\nu{\dot x
}^\mu-k({s}^{\mu\kappa}{F_\kappa}^\nu-s^{\nu\kappa}
{F_\kappa}^\mu), \label{eq2}
\end{equation}
supplemented by the following equations
\begin{eqnarray}
p_\mu s^{\mu\nu}=0, \label{eq111}
\end{eqnarray}
where $D$ represents covariant differentiation, ${\tau}$ is an
affine parameter along the worldline of the particle, ${\dot
x}^{\mu}=\frac{dx^{\mu}}{d\tau}$ is the 4-velocity of the
particle, $p^\mu$ is the components of its 4-momentum,
$s^{\mu\nu}$ is the particle's spin tensor, $q$ is the particle's
charge, $F^{\mu\nu}$ is the electromagnetic tensor, and
$k=\frac{qg}{2m}$ is a constant with $g$ being the particle's
gyromagnetic ratio and $m$ its mass, and
${R^\mu}_{\nu\alpha\beta}=\partial_\alpha\Gamma^\mu_{\nu\beta}-\partial_\beta
\Gamma^\mu_{\nu\alpha}+\Gamma^\mu_{\alpha\kappa}\Gamma^\kappa_{\nu\beta}
-\Gamma^\mu_{\beta\kappa}\Gamma^\kappa_{\nu\alpha}$. The
particle's spin may also be described by a four-vector
$$s^{\mu}=\frac{1}{2m\sqrt{-g}}{\varepsilon^{\mu}_{\,\,\,\,\nu\kappa\rho}
p^{\nu}s^{\kappa\rho}}.$$ It can be shown that these equations
lead to
\begin{equation}
\frac{1}{2}s_{\mu\nu}s^{\mu\nu}=s^2\label{e402}
\end{equation}
in which the spin $s$ of the particle is constant. We can also
deduce the following relation from the DS equations
\cite{souriau4}
\begin{equation}\label{e874}
p_\mu{\dot x}^\mu\frac{d}{d\tau}\left(p_\mu p^\mu\right)-p_\mu
p^\mu\frac{d}{d\tau}\left(\frac{qg}{2}F_{\mu\nu}s^{\mu\nu}\right)=0.
\end{equation}
We fix the gauge by
\begin{equation}\label{e875}
p_\mu{\dot x}^\mu-\frac{qg}{2m}F_{\mu\nu}s^{\mu\nu}=-m,
\end{equation}
which reduces to the Dixon's gauge introduced in \cite{dixon1} if
we let the electromagnetic field to be absent. Now it follows from
equation (\ref{e874}) that
\begin{equation}\label{e876}
p_\mu p^\mu-\frac{qg}{2}F_{\mu\nu}s^{\mu\nu}=-m^2
\end{equation}
is a constant of motion.

In the DS framework no direct equations of motion exist for ${\dot
x}^\mu$, but it can be shown that in the gauge (\ref{e875}), the
following relation results from equations
(\ref{eq1})-(\ref{eq111})
\begin{eqnarray}\label{e706}
{\dot x}^\mu&=&\frac{1}{m}p^\mu+\frac{qg}{2m(p_\mu
p^\mu)}\left(\frac{1}{2}s^{\alpha\beta}D^\mu F_{\alpha\beta}-p^\nu
s^{\mu\kappa}F_{\kappa\nu}\right) +\frac{s^{\mu\nu}f^\kappa
l_{\nu\kappa}}{1+ \frac{1}{2}s^{\kappa\rho}l_{\kappa\rho}}
\end{eqnarray}
where
$$l_{\nu\kappa}=\frac{1}{p_\mu p^\mu}\left(-\frac{1}{2}R_{\nu\kappa\alpha\beta}s^{\alpha\beta}
+qF_{\nu\kappa}\right).$$ The above equation reduces to the
equation given in \cite{fel} in the case of $F_{\mu\nu}=0$.
\section{Worldline deviations}
Let us now consider a one-parameter family of worldlines
$\{x^{\mu}({\tau,\lambda})\}$ describing the worldlines of charged
spinning particles of the same spin-to-mass ratios and the same
charge-to-mass ratios. In this family, worldlines are obtained
from a specific "fiducial" worldline via $x^\mu\rightarrow
x^\mu+\Delta\lambda n^\mu$ with $n^\mu=\frac{dx^\mu}{d\lambda}$.
We now define
\begin{equation}\label{eq4}
J^{\mu\nu}=\frac{Ds^{\mu\nu}}{D\lambda}\, ,\hspace{5mm}
j^{\mu}=\frac{Dp^\mu}{D\lambda}.
\end{equation}
Now by setting $x^{\mu}\rightarrow x^\mu+\Delta\lambda n^\mu$ in
equations (\ref{eq1})-(\ref{eq111}), and keeping only linear terms
of $\Delta\lambda$ in the resulting equations, and making use of
$$\frac{D{\dot x}^\mu}{D\lambda}=\frac{Dn^\mu}{D\tau},$$ we obtain the worldline deviation
equations, which read
\begin{eqnarray}
\frac{Dj^\mu}{D\tau}=&-&{R^\mu}_{\beta\alpha\kappa}{\dot x}
^{\kappa}n^{\alpha}p^{\beta}-\frac{1}{2}n^{\kappa}D_{\kappa}{R^\mu}_{\nu\lambda\rho}s^
{\lambda\rho}{\dot
x}^{\nu}-\frac{1}{2}{R^\mu}_{\nu\lambda\rho}J^{\lambda\rho}{\dot x
}^{\nu}\nonumber\\&-&\frac{1}{2}{R^\mu}_{\nu\lambda\rho}s^{\lambda\rho}
\frac{Dn^\nu}{D\tau}+q{F^\mu}_\beta\frac{Dn^{\beta}}{D\tau}+q\frac{D{F^\mu}_\alpha}{D\tau}{\dot
x}^\alpha \nonumber\\&+&\frac{qg}{4m}
J^{\kappa\rho}D^{\mu}F_{\kappa\rho}+\frac{qg}{4m}s^{\kappa\rho}
n^{\nu}D_{\nu}D^{\mu}F_{\kappa\rho}\label{eq9},\\
\frac{DJ^{\mu\nu}}{D\tau}=&&s^{\kappa[\mu}{R^{\nu]}}_{\kappa\alpha\beta}n^{\alpha}
{\dot x}^\beta+p^{[\mu}\frac{D}{D\tau}n^{\nu]}+j^{[\mu}{\dot
x}^{\nu]}\nonumber\\&-&\frac{qg}{2m}J^{[\mu\kappa}{{F_\kappa}^{\nu]}}
-\frac{qg}{2m}s^{[\mu\alpha}\frac{D}{D\lambda}{F_\alpha}^{\nu]}
,\label{eq10}
\end{eqnarray}
and
\begin{equation}\label{eq11}
s_{\mu\nu}j^{\nu}+ J_{\mu\nu}p^\nu=0,
\end{equation}
respectively. Here, $A^{[\mu}B^{\nu]}$ means $A^\mu B^\nu-A^\nu
B^\mu$. Similarly one can obtain from the equations (\ref{e402})
and (\ref{e876}) the following useful relations
\begin{eqnarray}
s_{\mu\nu}J^{\mu\nu}&=&0,\label{eq18}\\
p_{\mu}j^{\mu}-\frac{qg}{4}F_{\mu\nu}J^{\mu\nu}-\frac{qg}{4}s^{\mu\nu}
\frac{DF_{\mu\nu}}{D\lambda}&=&0,\label{eq19}
\end{eqnarray}
respectively. Also the relation (\ref{e875}) leads to
\begin{equation}\label{e719}
j_\mu{\dot
x}^\mu+p_\mu\frac{Dn^\mu}{D\tau}-\frac{qg}{2m}F_{\mu\nu}J^{\mu\nu}
-\frac{qg}{2m}s^{\mu\nu}\frac{DF_{\mu\nu}}{D\lambda}=0.
\end{equation}
To find the deviation $n^{\mu}$ itself, one should solve the above
equations for $j^{\mu},J^{\mu\nu}$ and find $n^{\mu}$ from these
indirectly. Having this in mind, a useful equation may be obtained
by starting from equation (\ref{e706}) and following the same
procedure described above. Thus we have
\begin{equation}\label{e877}
\frac{Dn^\mu(\tau)}{D\tau}=\frac{1}{m}j^\mu(\tau)+\frac{D}{D\lambda}
\left(\frac{r^\mu}{p_\mu p^\mu}\right)
\end{equation}
where
\begin{eqnarray*}
r^\mu=s^{\mu\nu}\left(-\frac{1}{2}R_{\nu\kappa\alpha\beta}{\dot
x}^\kappa s^{\alpha\beta}+qF_{\nu\kappa}{\dot
x}^\kappa+\frac{qg}{4m}s^{\alpha\beta}D_\nu
F_{\alpha\beta}\right)-\frac{qg}{2m}p^\nu
s^{\mu\kappa}F_{\kappa\nu}.
\end{eqnarray*}
\section{Motion in a gravitational wave}
Here we apply our results to the case of the motion of a charged
spinning particle in the space-time of a plane gravitational wave
when a uniform magnetic field is present. We take this magnetic
field to be in the same direction the wave propagates and the
gyromagnetic ratio as $g=2$.

The space-time metric is given by
\begin{equation}\label{eq20}
ds^2=-dudv-K(u,x,y)du^{2}+ dx^2+dy^2
\end{equation}
representing a gravitational wave propagating along the
$z$-direction. Here $(u,v)$ are the light-cone coordinates given
by $u=t-z$ and $v=t+z$ and $K(u,x,y)$ is given by
\begin{equation}\label{eq21}
K(u,x,y)=f(u)(x^2-y^2)
\end{equation}
in which $f(u)$ is an arbitrary function corresponding to the
linear polarization of the wave. We also take the non-vanishing
components of the electromagnetic tensor $F_{\mu\nu}$ as follows
\begin{equation}\label{eq25}
F_{34}=B
\end{equation}
which corresponds to a uniform magnetic field in the
$z$-direction. We label the coordinates $u,v,x,y$ with $1,2,3,4$
respectively.

With this metric and field, the DS equations admit the following
solution
\begin{eqnarray}
v^{\mu}&=&(1,1,0,0)\nonumber\\ p^{\mu}&=&(M,M,0,0)\label{eq23}\\
s^{34}&=&S,\hspace{3mm}s^{1\mu}=s^{2\mu}=0,\nonumber
\end{eqnarray}
where $M=\sqrt{m^2-qBS}$. This describes a particle sitting in the
origin of the coordinates with its spin directed along the
$z$-direction. We take the worldline of this particle,
$(\tau,\tau,0,0)$, as a fiducial worldline and calculate the
relative acceleration of nearby charged spinning particles and
also show that how approximate solutions to the DS equations may
be found in the vicinity of the above fiducial worldline. Now by
using the deviation equations of the previous section we reach at
\begin{equation}\label{e902}
n^1(\tau)=-n^2(\tau)=\alpha,
\end{equation}
and consequently
\begin{equation}
n^z(\tau)=-\alpha,\label{e902a}
\end{equation}
where $\alpha$ is a constant. This means that the particles gain
no relative velocity in the $z$-direction. For convenience we set
$\alpha=0$ hereafter. We also obtain
\begin{eqnarray}
\frac{dj^3(\tau)}{d\tau}&=&f(u)\left(J^{13}(\tau)-Mn^{3}(\tau)\right)+
qB\frac{dn^4(\tau)}{d\tau},\label{eq37}\\
\frac{dj^4(\tau)}{d\tau}&=&-f(u)\left(J^{14}(\tau)-Mn^{4}(\tau)\right)
-qB\frac{dn^3(\tau)}{d\tau},\label{eq38}\\
2Sj^4(\tau)&=&-M(J^{13}(\tau)+J^{23}(\tau)),\label{e998}\\
2Sj^3(\tau)&=&M(J^{14}(\tau)+J^{24}(\tau)),\label{e999}\\
J^{12}(\tau)&=&0,\label{e407}\\
J^{34}(\tau)&=&0,\label{e909b}\\
\frac{dJ^{13}(\tau)}{d\tau}&=&M\frac{dn^{3}(\tau)}{d\tau}-j^{3}(\tau)
+\omega J^{14}(\tau),\label{eq41}\\
\frac{dJ^{14}(\tau)}{d\tau}&=&M\frac{dn^{4}(\tau)}{d\tau}-j^{4}(\tau)
-\omega J^{13}(\tau),\label{eq40}\\
\frac{dJ^{23}(\tau)}{d\tau}&=&M\frac{dn^3(\tau)}{d\tau}-j^{3}(\tau)+
\omega J^{24}(\tau)-2Sf(u)n^4(\tau),\label{e414}\\
\frac{dJ^{24}(\tau)}{d\tau}&=&M\frac{dn^4(\tau)}{d\tau}-j^{4}(\tau)-\omega
J^{23}(\tau)-2Sf(u)n^3(\tau)\label{e415},\\
\frac{dn^3(\tau)}{d\tau}&=&\frac{1}{m}j^3(\tau)+\frac{S}{M^2-qBS}f(u)J^{14}(\tau)
,\label{e416}\\
\frac{dn^4(\tau)}{d\tau}&=&\frac{1}{m}j^4(\tau)+\frac{S}{M^2-qBS}f(u)J^{13}(\tau),
\label{e417}
\end{eqnarray}
where $\omega=\frac{qB}{m}$. The solutions to these equations give
$n^\mu$ and hence the relative accelerations of the particles near
the fiducial worldline and also approximate solutions of the DS
equations there. These equations are simplified in the interesting
situation where $M=0$, that is in the fine-tuned case of
$\frac{s}{m}=\frac{1}{\omega}$. In this case the above equations
result in
\begin{eqnarray}
j^3(\tau)&=&0=j^4(\tau),\label{eq37a}\\
J^{12}(\tau)&=&0=J^{34}(\tau),\label{e407a}\\
J^{13}(\tau)&=&J\sin(\omega\tau+\phi),\label{eq41a}\\
J^{14}(\tau)&=&J\cos(\omega\tau+\phi),\label{eq40a}\\
n^3(\tau)&=&-\frac{J}{qB}\int f(\tau)\cos(\omega\tau+\phi)d\tau,\label{e416a}\\
n^4(\tau)&=&-\frac{J}{qB}\int f(\tau)\sin(\omega\tau+\phi)d\tau,\label{e417a}\\
J^{23}(\tau)&=&-\cos(\omega\tau+\chi)\eta(\tau)-\sin(\omega\tau+\chi)\rho(\tau),\label{e414a}\\
J^{24}(\tau)&=&\sin(\omega\tau+\chi)\eta(\tau)-\cos(\omega\tau+\chi)\rho(\tau)\label{e415a}\\
\end{eqnarray}
with $J,\phi,\chi$ being constants and
\begin{eqnarray*}
\eta(\tau)&=&2S\int
(n^4(\tau)\cos(\omega\tau+\chi)-n^3(\tau)\sin(\omega\tau+\chi))f(\tau)d\tau,\\
\rho(\tau)&=&2S\int
(n^3(\tau)\cos(\omega\tau+\chi)+n^4(\tau)\sin(\omega\tau+\chi))f(\tau)d\tau.
\end{eqnarray*}
\section{Conclusions}
In this work we have studied worldline deviations of charged
spinning particles of the same spin-to-mass and the same
charge-to-mass ratios in the framework of the DS equations and
determined the effects of the spin-curvature, the
charge-electromagnetic field, and the spin-electromagnetic
couplings on the relative acceleration of nearby particles. The
equations we have found, reduce to those of \cite{moh2} if the
electromagnetic field is turned off.

The equation of geodesic deviation is usually considered in the
literature together with an extra equation, namely ${\dot x}_\mu
n^\mu=0$. This is because ${\dot x}_\mu n^\mu$ is constant along a
time-like geodesic. However, in the case of spinning particles,
the relation ${\dot x_\mu}{\dot x}^\mu=-1$ is no longer guaranteed
by the equations of motion and ${\dot x}_\mu n^\mu$ is not a
constant along the particle's worldlines. In the framework of the
DS equations, one may consider the equation (\ref{e719}) having
the same role the equation ${\dot x}_\mu n^\mu=0$ plays in the
case of geodesics.

It is possible to extend the above worldline deviation equations
systematically to equations of higher accuracy by keeping higher
order terms in $\Delta\lambda$. These extensions then may be used
to calculate the relative accelerations of particles of arbitrary
separations.

Another interesting application of our worldline deviation
equations is to generate approximate solutions of the DS equations
from a given solution. The application of these equations in the
study of the dynamics of charged spinning particles in some
interesting space-times would be reported in a future work.

{\bf Acknowledgement}. M. Mohseni would like to thank the Abdus
Salam ICTP where part of this work was done.

\end{document}